\documentclass[10pt,conference]{IEEEtran}
\usepackage{amssymb}
\usepackage{amsmath}
\usepackage{mathbbold}
\usepackage{amsfonts}
\usepackage{amsmath,graphicx,amssymb,algorithm, algorithmic}

\begin{document}

\title{Scheduling in Wireless Networks under Uncertainties: A Greedy Primal-Dual Approach}

\author{\authorblockN{Qiao Li}
\authorblockA{qiaoli@cmu.edu\\
Department of Electrical and Computer Engineering \\
Carnegie Mellon University \\
5000 Forbes Ave., Pittsburgh, PA 15213
} \and
\authorblockN{Rohit Negi}
\authorblockA{negi@ece.cmu.edu\\
Department of Electrical and Computer Engineering \\
Carnegie Mellon University\\
5000 Forbes Ave., Pittsburgh, PA 15213 }}

\maketitle

\newcommand{\diff}{{d\over dt}}
\newcommand{\bl}{\boldsymbol}
\newcommand{\ml}{\mathcal}
\newtheorem{theorem}{Theorem}
\newtheorem{lemma}{Lemma}
\newtheorem{assumption}{Assumption}

\begin{abstract}
This paper proposes a dynamic primal-dual type algorithm to solve
the optimal scheduling problem in wireless networks subject to
uncertain parameters, which are generated by stochastic network
processes such as random packet arrivals, channel fading, and node
mobilities. The algorithm is a generalization of the well-known
max-weight scheduling algorithm proposed by Tassiulas \emph{et al.},
where only queue length information is used for computing the
schedules when the arrival rates are uncertain. Using the technique
of fluid limits, sample path convergence of the algorithm to an
arbitrarily close to optimal solution is proved, under the
assumption that the Strong Law of Large Numbers (SLLN) applies to
the random processes which generate the uncertain parameters. The
performance of the algorithm is further verified by simulation
results. The method may potentially be applied to other applications
where dynamic algorithms for convex problems with uncertain
parameters are needed.
\end{abstract}

\section{Introduction}
\label{sec_intro}

Scheduling in wireless networks involves efficiently allocating
network resources among competing network users \emph{in the
presence of uncertainties}. These uncertainties may be either due to
unexpected events, such as link failures, or due to intricate
cross-layer interactions in wireless networks. For example, the
packet arrival rates may be unknown (e.g., \cite{Tassiulas92},
\cite{neely06}), which depend on upper layer dynamics such as
routing and congestion control protocols. As another example, the
wireless channel statistics may be also unknown (e.g.,
\cite{neely06}), since they depend on complex network events such as
channel fading, power control and node mobilities.

In the presence of such uncertain parameters, it may no longer be
optimal to use the static allocation approach (e.g.,
\cite{hajek88}), which produces periodic schedules by solving a
static underlying convex optimization problem (which usually has
exponential size) with estimated uncertain parameters. In
particular, if these uncertain parameters are slowly converging, or
time varying, the estimated parameters may fail to track the changes
in their true value, which often leads to suboptimal schedules.
Further, it may be impractical to estimate the uncertain parameters
for large wireless networks, as the number of the parameters may
grow fast (e.g., exponentially) with the size of the network. For a
simple illustration, consider a wireless network with $n$ links,
such that each link randomly switches on or off after certain random
number of time slots. In such case, a complete specification of the
network topology probabilities may require as large as $2^n-1$
parameters, which quickly becomes impossible to estimate as $n$
grows.

On the other hand, online algorithms (such as \cite{Tassiulas92},
\cite{neely06}, \cite{georgiadis}, \cite{stolyar05},
\cite{eryilmaz06}) are more robust to the changes to the uncertain
parameters (such as arrival rates), since they use queue length
(which can also be interpreted as prices \cite{kelly98},
\cite{shakkottai07}) information for scheduling in each time slot.
For example, it has been shown that \cite{georgiadis} such
algorithms can achieve network stability even if the ``instantaneous
rates'' of the traffic vary arbitrarily inside the network capacity
region. Further, compared to the estimation based approach, these
online algorithms may be more scalable to the network size, in the
sense that the dimension of the queue length vector corresponds to
the number of constraints (such as rate constraint for each link),
which usually grows slowly, whereas the number of uncertain
parameters can grow very fast (e.g., exponentially).

In this paper we solve a general class of optimal wireless network
scheduling problems with uncertain parameters, whose underlying
static problem is described by the convex optimization problem
\textsf{OPT} in Section \ref{sec_model_formulation}. Essentially, we
require that the structures of the convex objective functions and
convex constraint functions are known, except the values of the
uncertain parameters. These parameters will be generated by certain
stochastic processes and observed by the network gradually over time
slots. We propose a greedy primal-dual dynamic algorithm (Algorithm
\ref{alg} in Section \ref{sec_algorithm}) to achieve the optimal
scheduling asymptotically. Using the novel technique of fluid limits
\cite{Dai00}, optimality can be guaranteed under the assumption that
all the network processes generating the uncertain parameters
satisfy SLLN (see details in Section \ref{sec_model_formulation}).
Note that this assumption is quite mild, since we can guarantee
optimality \emph{as long as these processes converge, no matter how
slowly the convergence happen}. Thus, intuitively, our algorithm can
automatically track the convergence of these processes and correct
the mistakes which are made within any finite time history.

Our algorithm is a generalization of the well-known max-weight
algorithm \cite{Tassiulas92}, which was shown to be throughput
optimal for i.i.d arrival processes. Our algorithm is related to,
but different from the utility-optimal scheduling algorithm by Neely
\cite{neely06}, which achieves the optimal scheduling by cleverly
transforming the problem into optimizing the time-average of the
utility (and constraint) functions, to which a dual-type algorithm
applies. Our algorithm is also different from the primal-dual
algorithm by Neely \cite{neely10} since we use a scaled queue length
(by $1/t$) in the scheduling, which corresponds to the approximated
gradient. Stolyar \cite{stolyar05} also proposed a primal-dual type
scheduling algorithm, and proved its optimality using a fluid limit
obtained from a different scaling. Since the fluid scaling in
\cite{stolyar05} is taken over different systems, it is hard to
relate the optimality in the fluid limits to the one in the original
system. Finally, our algorithm can be used as a MAC layer solution
for the general framework of cross-layer optimization problem (e.g.,
\cite{georgiadis}, \cite{eryilmaz06}, \cite{eryilmaz07},
\cite{shakkottai07}) for wireless networks.

The organization of the following sections is as follows: In Section
\ref{sec_model_formulation} we describe the queueing network model
as well as the optimization problem \textsf{OPT}, and in Section
\ref{sec_algorithm} we describe the scheduling algorithm. Section
\ref{sec_analysis} proves the optimality of the algorithm, Section
\ref{sec_simulation} illustrate the algorithm performance in
simulation, and finally Section \ref{sec_conclusion} concludes this
paper.

\section{Network Model and Problem Formulation}
\label{sec_model_formulation}

In this section we describe the queueing network and propose the
optimization problem \textsf{OPT}. We first introduce the queueing
network model.

\subsection{Queueing Network}

We consider the scheduling problem at the medium access (MAC) layer
of a multi-hop wireless network, where the network is modeled as a
set of $n$ links. We assume a time-slotted network model, and in
each time slot $t$, the network is in one of the following states:
$\ml M\triangleq\{1, 2, \ldots, M\}$. The network state can be used
to model network topology, channel fading, and user mobility, etc.
We further assume that these network states can be measured by the
user nodes\footnote{Note that although the network states are
global, they often allow (approximate) decompositions (e.g.,
\cite{neely06}) according to either the geographic structure or
channel orthogonality, in which case one only needs to measure local
network states.}, which are assumed to be equipped with channel
monitoring devices. We associate each network state $m\in \ml M$
with a finite set of resource allocation modes
$\Xi^{(m)}=\{\xi^{(m)}_1, \xi^{(m)}_2, \ldots\}$, where each mode
$\xi^{(m)}_k\in \Xi^{(m)}$ corresponds to a configuration of network
resource allocation, such as carrier and frequency selection in OFDM
systems, spreading codes choice in CDMA systems and time slots
assignment in TDMA systems.

Denote $\bl y$ as the uncertain parameters, which are generated by
the stochastic process $\bl Y(t)$, which is a cumulative vector
process whose time average converges to $\bl y$. Specifically, the
assumptions on $\bl Y(t)$ are: 1) it is subject to SLLN, i.e., with
probability 1 (w.p.1),
\begin{equation}
\lim_{t\rightarrow\infty}\bl Y(t)/t = \bl y
\end{equation}
and 2) it has uniformly bounded increment in each time slot:
\begin{equation}
\|\bl Y(t)-\bl Y(t-1)\|\leq Y_{\max}, \forall t>0
\end{equation}
where $Y_{\max}>0$ is a finite constant. For specific examples,
consider the cumulative network state process $M_m(t)$:
\begin{equation}
S_m(t)=\sum_{\tau=1}^t\mathbbold{1}_{\{m(\tau)=m\}}
\end{equation}
where $m(t)$ is the network state at time slot $t$, and
$\mathbbold{1}_{\{\cdot\}}$ is the indicator function, i.e.,
$\mathbbold{1}_{\{\text{true}\}}=1$ and
$\mathbbold{1}_{\{\text{false}\}}=0$. Thus, SLLN implies that w.p.1,
\begin{equation*}
\lim_{t\rightarrow\infty}{1\over t}
S_m(t)=\lim_{t\rightarrow\infty}{1\over
t}\sum_{\tau=1}^t\mathbbold{1}_{\{m(\tau)=m\}}=\pi^{(m)},\ \forall
m\in\mathcal{M}
\end{equation*}
As another example, consider the external packet arrival process
$\bl A(t)$, which is a $n\times 1$ vector representing the
cumulative external packet arrivals during the first $t$ time slots.
Similarly, SLLN implies that w.p.1,
\begin{equation}
\lim_{t\rightarrow\infty}{\bl A}(t)/t={\bl a},\ \textrm{w.p.1}
\end{equation}
Further, we require that the maximum packet arrivals in any time
slot are uniformly bounded:
\begin{equation}
\|\bl A(t)-\bl A(t-1)\|\leq A_{\max},\ \forall t>0
\end{equation}
where $A_{\max}>0$ is a finite constant.

We finally describe the queueing system model. The queuing dynamics
of the network is modeled as follows:
\begin{equation}\label{eqn_qt}
{\bl Q}(t)={\bl Q}(0)+{\bl A}(t)-R(m(t)){\bl D}(t)
\end{equation}
where ${\bl Q}(t)$ is the queue length vector at time slot $t$, and
$R(m(t))$ is the $n\times n$ routing matrix, such that
$R_{ii}(m(t))=1$, and $R_{ij}(m(t))=-1$ only if link $i$ serves as
the next hop for link $j$ at time slot $t$, as specified by certain
routing protocols, otherwise $R_{ij}(m(t))=0$. Note that the routing
matrix $R(m(t))$ is a function of the network state, and therefore
SLLN implies
\begin{equation}
\lim_{t\rightarrow\infty}{1\over t}\sum_{\tau=1}^t
R(m(\tau))=\sum_{m\in\ml M}\pi^{(m)}R(m)
\end{equation}
${\bl D}(t)$ is a $n\times 1$ vector representing the cumulative
packet departures during the first $t$ time slots, which are
determined by the resource allocation modes as specified by the
scheduler in each time slot. Specifically, at each time slot $t$
with network state $m$, if the scheduler chooses a resource
allocation mode $\xi_k^{(m)}$, there is an associated departure
vector $\bl G^{(m)}_k$, whose each entry ${G}^{(m)}_{ki}$
corresponds to the number of packets transmitted successfully by
link $i$. Note that the choice of resource allocation mode
$\xi_k^{(m)}$ is subject to the constraint that $\bl Q(t)-\bl
G^{(m)}_k\succeq \bl 0$, so that the queue lengths never become
negative. Note that this constraint can be easily satisfied in
general systems. For example, if the allocation mode $\xi_k^{(m)}$
corresponds to independent sets of the interference graph (see, for
example, \cite{eryilmaz06}), one can simply transmit the subset of
links with nonempty queues, which are still independent. In a
compact form, we can express the departure process as
\begin{equation}\label{eqn_dt}
{\bl D}(t)=\sum_{m\in{\ml M}}G^{(m)}{\bl T}^{(m)}(t),
\end{equation}
where $G^{(m)}$ is a matrix whose columns are $\bl G^{(m)}_k$, and
${\bl T}^{(m)}(t)$ is a vector whose each entry ${ T}^{(m)}_k(t)$
corresponds to the number of time slots that resource allocation
mode $\xi^{(m)}_k$ is chosen during the first $t$ time slots.

A basic requirement on the scheduler is that it should achieve rate
stability \cite{Dai00}, i.e.,
\begin{equation}
\lim_{t\rightarrow\infty}\bl D(t)/t=\bl a
\end{equation}
so that the departure rate of each link is equal to the arrival
rate, as required by the underlying static optimization problem
\textsf{OPT}, which we formulate in the next subsection.

\subsection{Optimization Problem}

In this section we introduce the optimization problem \textsf{OPT},
which is implicitly solved by the optimal schedulers. The problem
\textsf{OPT} is as follows \textsf{OPT}:
\begin{eqnarray}
\textsf{OPT: }\min_{{\boldsymbol x}}&&f({\bl x}; {\bl y})\label{eqn_cost}\\
\textrm{s.t.}&&\bl h({\bl x}; {\bl y})\preceq{\bl 0}\label{eqn_cvx_const}\\
&&{\boldsymbol x}^{(m)}\succeq 0,\ {\boldsymbol 1}^T{\boldsymbol
x}^{(m)}=1,\ \forall m\in\mathcal{M}\label{eqn_simplex_const}
\end{eqnarray}
In the above formulation, $\bl x^{(m)}$ as a resource allocation
vector when the network state is $m$. That is, each entry
$x_k^{(m)}$ is the asymptotic time fraction (assuming the limit
exists for now) that resource allocation mode $\xi^{(m)}_k$ is
chosen, during the time slots where the network state is $m$. Thus,
$\bl x^{(m)}$ is subject to the simplex constraint
(\ref{eqn_simplex_const}). $\bl x=(\bl x^{(1)}, \bl x^{(2)}, \ldots,
\bl x^{(M)})$ is a big vector representing the total resource
allocation vector as specified by the scheduler. $f({\bl x}; {\bl
y})$ is a general convex cost function of variable $\bl x$, and $\bl
h({\bl x}; {\bl y})$ is a vector of general convex constraint
functions of variable $\bl x$. The additional parameter ${\bl y}$
represents the uncertain parameters, which is valid under the
assumption that the corresponding processes are subject to SLLN.
Finally, we assume that both $f({\bl x}; {\bl y})$ and $h({\bl x};
{\bl y})$ are continuously differentiable as functions of variables
$(\bl x, \bl y)$.

The formulation of \textsf{OPT} is quite general, which can be used
to model various applications in the literature. For example, if we
want to minimize the total transmission power, we can choose $\bl
y=(\bl \pi, \bl p)$, and choose the cost function as follows
\begin{equation}\label{eqn_power_cost}
f({\bl x}; {\bl y})=\sum_{m\in\ml M} \pi^{(m)}\bl p^{(m)T}\bl
x^{(m)}
\end{equation}
where $\pi^{(m)}$ is the time fraction that the network state is
$m$, and $\bl p^{(m)}$ is a power vector where each entry
$p^{(m)}_k$ corresponds to the power consumption when resource
allocation mode $\xi^{(m)}_k$ is chosen at network state $m$. Thus,
the cost function in (\ref{eqn_power_cost}) can be interpreted as
the average power consumption by the scheduler. Note that we can
also encode the power constraint into $\bl h({\bl x}, {\bl y})$ by
choosing $\bl y=(\bl \pi, \bl p)$ and then choosing
\begin{equation}\label{eqn_power_const}
\bl h(\bl x; \bl y)=\sum_{m\in\ml M}\pi^{(m)} P^{{(m)}}\bl
x^{(m)}-\bl p
\end{equation}
where $ P^{{(m)}}$ is a diagonal matrix where each diagonal entry
$P^{(m)}_{kk}$ corresponds to the power consumption when resource
allocation mode $\xi_k^{(m)}$ is chosen when the network state is
$m$, and $\bl p$ is the power constraint vector. In this case,
(\ref{eqn_power_const}) is equivalent to requiring a constraint of
$\bl p$ on the average transmission power. In order to encode the
network stability constraint, we can choose $\bl y=(\bl \pi, \bl a)$
and then choose
\begin{equation}\label{eqn_stab_const}
\bl h({\bl x; \bl y})=\bl a-\sum_{m\in\ml M}\pi^{(m)} R(m)
G^{{(m)}}\bl x^{(m)}
\end{equation}
Thus, (\ref{eqn_stab_const}) requires that the average external and
internal arrivals should be less than the average departures, in
which case the network is rate stable.

\section{Scheduling Algorithm}
\label{sec_algorithm}

In this section we will describe the algorithm to solve
\textsf{OPT}. As a standard approach in solving constrained convex
optimization problems \cite{bertsekas99}, we transform \textsf{OPT}
into another static ``penalized'' problem, \textsf{PEN}, to which
our scheduling algorithm can directly apply. Based on this, we then
introduce the scheduling algorithm which solves \textsf{PEN} and,
therefore, also solves \textsf{OPT}.

\subsection{Transformed Problem}

Assuming that \textsf{OPT} is strictly feasible, we first change the
constraints in (\ref{eqn_cvx_const}) as follows
\begin{eqnarray}\label{eqn_equal_const}
&&\bl h({\bl x}; \bl y)+{\bl z}={\boldsymbol 0},\ \epsilon{\bl
1}\preceq {\bl z}\preceq{z}_{\max}\bl 1
\end{eqnarray}
where $\epsilon>0$ is a small scalar, and ${z}_{\max}>0$ is a
sufficiently large constant such that the inequality and equality
constraints are equivalent. Denote $f^\star_\epsilon$ as the optimal
cost when the constraint is changed to (\ref{eqn_equal_const}).
Thus, the optimal value of \textsf{OPT} is $f^\star_0$. We have the
following sensitivity lemma stating that $f^\star_\epsilon$ is a
good approximation of $f^\star_0$ with sufficiently small
$\epsilon$.

\begin{lemma}[\cite{bertsekas99}]\label{lem_approx_1}
Denote $\bl \lambda_0$ and $\bl \lambda_\epsilon$ as two Lagrangian
multipliers for $f^\star_0$ and $f^\star_\epsilon$, respectively. We
have
\begin{equation}
|f^\star_0-f^\star_\epsilon|\leq\epsilon\max(\|\bl
\lambda_\epsilon\|_1, \|\bl \lambda_0\|_1)
\end{equation}
\end{lemma}

We next define the transformed problem as follows:
\begin{eqnarray*}
\textsf{PEN: }\min_{{\bl x, \bl z}}&&g({\bl x}, {\bl z}; \bl y)=f({\bl x}; \bl y)+\beta p({\bl x}, {\bl z}; \bl y)\\
\textrm{s.t. }&&\epsilon{\bl 1}\preceq {\bl z}\preceq{
z}_{\max}\bl 1\\
&&{\boldsymbol x}^{(m)}\succeq 0,\ {\boldsymbol 1}^T{\boldsymbol
x}^{(m)}=1,\ \forall m\in\mathcal{M}
\end{eqnarray*}
where $\beta$ is a large constant to control approximation accuracy,
and $p({\bl x}, {\bl z}; \bl y)$ corresponds to the penalty term,
which corresponds to various standard penalty functions
\cite{bertsekas99}, e.g.,
\begin{equation}\label{eqn_penalty_example}
p({\bl x}, {\bl z}; \bl y)={1\over \alpha}\|\bl h({\bl x}; \bl
y)+{\bl z}\|^\alpha
\end{equation}
for $\alpha>1$. In particular, the standard Lyapunov drift analysis
(e.g., \cite{Tassiulas92}, \cite{neely06}, \cite{georgiadis},
\cite{eryilmaz06}) corresponds to the case $\alpha=2$.

Denote $({\bl x}^\star_{\textsf{p}}, {\bl z}^\star_{\textsf{p}})$ as
a solution of \textsf{PEN}. We have the following result holds:

\begin{lemma}\label{lem_approx_2}
$f({\bl x}^\star_{\textsf{p}}; \bl y)\leq f^\star_\epsilon$.
\end{lemma}

\begin{IEEEproof}
Denote $(\bl x^\star_{\textsf{o}}, \bl z^\star_{\textsf{o}})$ as a
solution of \textsf{OPT} with constraint in (\ref{eqn_cvx_const})
replaced by (\ref{eqn_equal_const}). We have
\begin{eqnarray*}
f({\bl x}^\star_{\textsf{p}}; \bl y)&\leq& f({\bl
x}^\star_{\textsf{p}};
\bl y)+\beta p({\bl x}^\star_{\textsf{p}}, {\bl z}^\star_{\textsf{p}}; \bl y)\\
&\stackrel{(a)}{\leq} & f({\bl x}^\star_{\textsf{o}}; \bl
y)+\beta p({\bl x}^\star_{\textsf{o}}, {\bl z}^\star_{\textsf{o}}; \bl y)\\
&\stackrel{(b)}{=}&f^{\star}_\epsilon
\end{eqnarray*}
where $(a)$ is because $(\bl x^\star_{\textsf{p}}, \bl
z^\star_{\textsf{p}})$ solves \textsf{PEN}, and $(b)$ is because
$(\bl x^\star_{\textsf{o}}, \bl z^\star_{\textsf{o}})$ satisfy the
constraint (\ref{eqn_equal_const}). Thus, the claim holds.
\end{IEEEproof}
In the following we will focus on solving \textsf{PEN}, since Lemma
\ref{lem_approx_1} and Lemma \ref{lem_approx_2} guarantee that
\textsf{PEN} achieve an objective function value which is
arbitrarily close to the optimal in \textsf{OPT}. We next describe
the scheduling algorithm.

\subsection{Algorithm Description}

The problems \textsf{OPT} and \textsf{PEN} are static. On the other
hand, the network is dynamic, and must be described by time series.
Therefore, before describing the algorithm, we need to define
dynamic counterparts of the static variables $\bl x, \bl y$ and $\bl
z$.  We first define empirical resource allocation variable
\begin{equation}\label{eqn_x}
{\bl x}^{(m)}(t)={{\bl T}^{(m)}(t)\over {\bl 1}^T{\bl T}^{(m)}(t)},\
\forall m\in\ml{M}
\end{equation}
i.e., each entry ${x}_k^{(m)}(t)$ corresponds to the time fraction
that resource allocation mode ${\xi}^{(m)}_k$ is chosen during the
first $t$ time slots, when the network state is $m$. Note that we
have
\begin{equation}
{\boldsymbol x}^{(m)}(t)\succeq 0,\ {\boldsymbol 1}^T{\boldsymbol
x}^{(m)}(t)=1,\ \forall m\in\mathcal{M}
\end{equation}
Thus, ${\bl x}^{(m)}(t)$ can be interpreted as the empirical value
of ${\bl x}^{(m)}$ which is defined in \textsf{PEN}. Similarly, we
denote the empirical value of the uncertain parameter $\bl y$ as
\begin{equation}\label{eqn_y}
\bl y(t)=\bl Y(t)/t
\end{equation}
i.e., $\bl y(t)$ is formed by directly taking the average of the
process $\bl Y(t)$. Further, define the empirical value of $\bl z$
as
\begin{equation}\label{eqn_z}
\bl z(t)=\bl Z(t)/t
\end{equation}
where the cumulative process $\bl Z(t)$ is defined by
\begin{equation}
\bl Z(t)=\sum_{\tau=1}^t \bl u(\tau)
\end{equation}
and $\bl u(\tau)$ is computed by the scheduler in Algorithm
\ref{alg}.

Finally, we introduce some notations. Denote $\nabla_m$ and
$\nabla_{\bl z}$ as the gradient operator with respect to variables
$\bl x^{(m)}$ and $\bl z$, respectively. Further, with an abuse of
notation, we use the following abbreviated notations:
\begin{eqnarray*}
f(t)&\triangleq&f(\bl x(t);\bl y(t))\\
p(t)&\triangleq&p(\bl x(t), \bl z(t);\bl y(t))\\
g(t)&\triangleq&g(\bl x(t), \bl z(t);\bl y(t))=f(t)+\beta p(t)
\end{eqnarray*}

\begin{algorithm}
\caption{Optimal Scheduling} \label{alg}
\begin{algorithmic}
\STATE {\bf Step 1.} At each time slot $t$ with network state $m$,
choose allocation mode $\xi^{(m)}_k$, where
\begin{eqnarray}\label{eqn_mink}
k\in\arg\min_{j}\big(\nabla_m f(t)+\beta\nabla_m p(t)\big)_j
\end{eqnarray}

\STATE {\bf Step 2.} Choose variable $\bl u(t)$ such that
\begin{equation}
u_i(t)= \left\{ \begin{array}{ll}
         \epsilon & \textrm{ if } (\nabla_{\bl z} p(t))_i\geq 0\\
                  z_{\max}& \textrm{ else }
                  \end{array} \right.
\end{equation}
and update variables $\bl x(t)$, $\bl y(t)$ and $\bl z(t)$
accordingly.
\end{algorithmic}
\end{algorithm}

The algorithm is described as in Algorithm \ref{alg}. Essentially,
the algorithm updates the variables $\bl x(t)$ and $\bl z(t)$ by
computing descent directions $\bl v^{(m)}(t)$ and $\bl u(t)$ in {\bf
Step 1} and {\bf Step 2}, respectively, where $\bl v^{(m)}(t)$ is an
all-zero vector except an one at the $k$-th entry. Further note that
constraint $\bl Q(t)\succeq \bl G^{(m)}_k$ can be satisfied
implicitly with regular cost and penalty functions, i.e., assuming
the cost for transmitting a set of links is always no smaller than
that of transmitting any of its subsets.

From the definition of $\bl x(t)$ and $\bl z(t)$, these processes
are naturally updated as follows
\begin{eqnarray*}
\bl x^{(l)}(t)&=&\bl x^{(l)}(t-1),\ l\neq m\\
\bl x^{(m)}(t)&=&\bl x^{(m)}(t-1)+{1\over \bl 1^T \bl
T^{(m)}(t)}(\bl v^{(m)}(t)-\bl
x^{(m)}(t-1))\\
\bl z(t)&=&\bl z(t-1)+{1\over t}(\bl u(t)-\bl z(t-1))
\end{eqnarray*}
Thus, Algorithm \ref{alg} can be viewed as a stochastic gradient
algorithm for \textsf{PEN}, where the randomness comes from the time
varying functions $f(t)$ and $p(t)$, which are subject to the
changes in uncertain parameters $\bl y(t)$.

The optimization of (\ref{eqn_mink}) requires tracking the variables
$\bl x(t)$ and $\bl y(t)$, in general. However, in applications the
structure of the cost function $f(\bl x; \bl y)$ and penalty
function $p(\bl x; \bl y)$ often allows a much simpler computation.
For example, in the important case of optimal power scheduling,
where the cost function is formulated as (\ref{eqn_power_cost}) and
the constraint is as (\ref{eqn_stab_const}) with the typical value
$\alpha=2$, we have
\begin{eqnarray*}
\nabla_m f(t)+\beta\nabla_m p(t)=\pi^{(m)}(t)(\bl
p^{(m)}+{\beta\over t} (R(m)G^{(m)})^T\bl Q(t))
\end{eqnarray*}
where $\pi^{(m)}(t)={\bl 1^T \bl T^{(m)}(t)\over \sum_{l\in\ml M}\bl
1^T \bl T^{(l)}(t)}$ is the empirical time fraction of network state
$m$. Thus, the optimization in (\ref{eqn_mink}) essentially only
requires the queue length information (note that $\pi_m(t)$ becomes
an irrelevant scaling factor in the optimization). In particular, if
we are only interested in the rate stability, i.e., setting the
objective function as $f(\bl x; \bl y)=0$, the optimization in
(\ref{eqn_mink}) is equivalent to
\begin{eqnarray}
k\in\arg\min_j \big(G^{(m)T}(R(m)^T\bl Q(t))\big)_j
\end{eqnarray}
which is the same as the max-weight back-pressure algorithm proposed
by \cite{Tassiulas92}. We finally conclude this section by the
following lemma, which formally shows, essentially, the descent
property of Algorithm \ref{alg}.

\begin{lemma}\label{lem_cond_grad} The following properties hold for
Algorithm \ref{alg}:
\begin{enumerate}
\item[1)] $\bl v^{(m)}(t)$ solves the following problem
\begin{eqnarray}
\textsf{GRAD-X: }\min_{{\bl v}^{(m)}}&&\nabla_m g(t)^T \bl v^{(m)}\label{eqn_lp_v}\\
\textrm{s.t. }&&\bl v^{(m)}\succeq 0,\ {\boldsymbol 1}^T{\bl
v}^{(m)}=1\nonumber
\end{eqnarray}
\item[2)] $\bl u(t)$ solves the following problem
\begin{eqnarray}
\textsf{GRAD-Z: }\min_{{\bl u}}&&\nabla_{\bl z} g(t)^T \bl u\label{eqn_lp_u}\\
\textrm{s.t. }&&\epsilon\bl 1\preceq\bl u\preceq z_{\max}\bl
1\nonumber
\end{eqnarray}
\end{enumerate}
\end{lemma}
Thus, the variables $(\bl v, \bl u)$ computed by Algorithm \ref{alg}
can be interpreted as the points in the feasible region of
\textsf{PEN} which achieves the minimum inner product with the
corresponding (stochastic) gradients.
\begin{IEEEproof}
For 1), note that \textsf{GRAD-X} is a Linear Programming (LP)
problem over a simplex, and therefore the solution can be obtained
at a vertex \cite{bertsekas99} with the minimum directional
derivative. For 2), note that \textsf{GRAD-Z} is an LP over a
hypercube, and therefore the solution is obtained at the boundary
points. Thus the claim follows by noting that $\nabla_{\bl z}
g(t)=\nabla_{\bl z} p(t)$, since $f(\cdot)$ is not a function of
$\bl z$.
\end{IEEEproof}

\section{Optimality Proof}
\label{sec_analysis}

In this section we will prove the optimality of Algorithm \ref{alg}.
There are two issues to consider: 1) We need to show that Algorithm
\ref{alg} achieves the optimality of \textsf{OPT} asymptotically,
and 2) We need to show that Algorithm \ref{alg} is feasible for
\textsf{OPT}, i.e., constraint (\ref{eqn_cvx_const}) can not be
violated. We first briefly introduce fluid limits, which serves as
the key technique for the optimality proof.

\subsection{Fluid Limits}

We extend the domain of all processes to continuous time by linear
interpolation, and define the fluid scaling of a function $l(t)$ as
\begin{equation}
l^r(t)=l(rt)/r
\end{equation}
where $l$ can be functions ${\boldsymbol T}, {\bl Y}$ and $\bl Z$.
It can be shown that these scaled functions are uniformly
Lipschitz-continuous. Thus, according to the Arzela-Ascoli Theorem
\cite{royden}, any sequence of functions which is indexed by
$\{r_n\}_{n=1}^\infty$, i.e., $({\boldsymbol T}^{r_n}, {\boldsymbol
Y}^{r_n}, \bl Z^{r_n})$, contains a subsequence
$\{r_{n_k}\}_{k=1}^\infty$ which converges uniformly on compact sets
to a set of absolutely continuous functions (and, therefore,
differentiable almost everywhere \cite{royden}) $(\bar{\boldsymbol
T}, \bar{\boldsymbol Y}, \bar{\bl Z})$. Define any such limit as a
fluid limit. (Note that fluid limits are denoted by a bar.) We next
state some properties of the fluid limits.

\begin{lemma}
The processes in any fluid limit satisfies the following: For any
$t>0$, we have w.p.1,
\begin{equation}\label{eqn_Zt_bound}
\epsilon \bl 1\preceq \bar{\bl Z}(t)/t\preceq z_{\max}\bl 1
\end{equation}
and the following properties hold w.p.1: For all $t\geq 0$
\begin{eqnarray}
\bar{\boldsymbol Y}(t)&=&{\boldsymbol y}t\label{eqn_y_bar}\\
{\bl 1}^T\bar{\bl T}^{(m)}(t)&=&\pi^{(m)}t\quad\forall
m\in\mathcal{M}\label{eqn_T_pi}
\end{eqnarray}
\end{lemma}
\begin{IEEEproof}
(\ref{eqn_Zt_bound}) follows from Algorithm \ref{alg}, where each
$u_i(t)$ is chosen between $\epsilon$ and $z_{\max}$.
(\ref{eqn_y_bar}) and (\ref{eqn_T_pi}) follows directly from the
(functional) SLLN.
\end{IEEEproof}

We next define the resource allocation variables and auxiliary
variables in fluid limit as follows (one can compare with
(\ref{eqn_x}) and (\ref{eqn_z}) for similarities)
\begin{eqnarray}
\bar{\boldsymbol x}^{(m)}(t)&=&{1\over{\pi}^{(m)}t}\bar{\boldsymbol
T}^{(m)}(t)\label{eqn_x_bar}\\
\bar{\bl z}(t)&=&{\bar{\bl Z}(t)/t}
\end{eqnarray}
Similarly, define the following variables as the counter parts of
$\bl v^{(m)}(t)$ and $\bl u(t)$ in Algorithm \ref{alg}:
\begin{eqnarray}
\bar{\boldsymbol v}^{(m)}(t)&=&\dot{\bar{\boldsymbol
T}}^{(m)}(t)/\pi^{(m)}\label{eqn_vbar}\\
\bar{\boldsymbol u}(t)&=&\dot{\bar{\boldsymbol
Z}}(t)\label{eqn_ubar}
\end{eqnarray}
We have the following lemma holds, which states that both $(\bar{\bl
x}(t), \bar{\bl z}(t))$ and $(\bar{\bl v}(t), \bar{\bl u}(t))$ are
feasible points for \textsf{PEN}.
\begin{lemma}\label{lem_feasible}
For any fluid limit and $t>0$, we have
\begin{enumerate}
\item[1)] $(\bar{\bl
x}(t), \bar{\bl z}(t))$ is feasible for \textsf{PEN}:
\begin{eqnarray}
&&\epsilon\bl 1\preceq \bar{\bl z}(t)\preceq z_{\max}\bl 1\label{eqn_zbar_bound}\\
&&{\boldsymbol 1}^T\bar{\boldsymbol x}^{(m)}(t)=1, \bar{\boldsymbol
x}^{(m)}(t)\succeq {\boldsymbol 0}, \forall m\in \ml
M\label{eqn_xbar_simplex}
\end{eqnarray}

\item[2)] $(\bar{\bl v}(t), \bar{\bl u}(t))$ is also feasible for \textsf{PEN}:
\begin{eqnarray}
&&\epsilon\bl 1\preceq \bar{\bl u}(t)\preceq z_{\max}\bl 1\\
&&{\boldsymbol 1}^T\bar{\boldsymbol v}^{(m)}(t)=1, \bar{\boldsymbol
v}^{(m)}(t)\succeq {\boldsymbol 0}, \forall m\in \ml
M\label{eqn_vbar_simplex}
\end{eqnarray}

\item[3)] The derivatives of $\bar{\boldsymbol x}^{(m)}(t)$ and $\bar{\bl
z}(t)$ are
\begin{eqnarray}\label{eqn_diff_xbar}
\dot{\bar{\boldsymbol x}}^{(m)}(t)&=&({\bar{\boldsymbol
v}}^{(m)}(t)-{\bar{\boldsymbol x}}^{(m)}(t))/t\\
\dot{\bar{\boldsymbol z}}(t)&=&({\bar{\boldsymbol
u}}(t)-{\bar{\boldsymbol z}}(t))/t
\end{eqnarray}
\end{enumerate}
\end{lemma}
\begin{IEEEproof}
For 1), (\ref{eqn_zbar_bound}) follows from applying
(\ref{eqn_Zt_bound}) to the definitions of $\bar{\bl z}(t)$, and
(\ref{eqn_xbar_simplex}) follows from applying (\ref{eqn_T_pi}) to
the definition of $\bar{\boldsymbol x}^{(m)}(t)$. Similarly we can
prove 2), by noting that $\epsilon\tau\bl 1\preceq \bar{\bl
Z}(t+\tau)-\bar{\bl Z}(t)\preceq z_{\max}\tau\bl 1$ for any $t\geq0$
and $\tau>0$. 3) follows from direct calculation.
\end{IEEEproof}

We are now ready to prove the optimality of Algorithm \ref{alg}.

\subsection{Optimality Proof}

For the ease of presentation, we use
\begin{equation}
\bar{g}(t)\triangleq g(\bar{\bl x}(t), \bar{\bl z}(t); \bar{\bl
y}(t))
\end{equation}
as a short-hand notation (note that they are the functions in fluid
limits), with an abuse of notation. We next establish the following
key technical lemma, which, essentially, extends the optimality
property in Lemma \ref{lem_cond_grad} to the fluid limits.

\begin{lemma}\label{lem_fluid_grad}
Let a fluid limit $(\bar{\boldsymbol T}, \bar{\bl Y}, \bar{\bl Z})$
and $m\in\ml M, t> 0$ be given. The following properties hold:
\begin{enumerate}
\item[1)] $\bar{\bl v}^{(m)}(t)$ solves the following problem
\begin{eqnarray}
\textsf{GRAD-XBAR: }\min_{\bar{\bl v}^{(m)}}&&\nabla_m \bar{g}(t)^T \bar{\bl v}^{(m)}\label{eqn_lp_v}\\
\textrm{s.t. }&&\bar{\bl v}^{(m)}\succeq 0,\ {\boldsymbol
1}^T\bar{\bl v}^{(m)}=1\nonumber
\end{eqnarray}
\item[2)] $\bar{\bl u}(t)$ solves the following problem
\begin{eqnarray}
\textsf{GRAD-ZBAR: } \min_{\bar{\bl u}}&&\nabla_{\bl z} \bar{g}(t)^T \bar{\bl u}\label{eqn_lp_u}\\
\textrm{s.t. }&&\epsilon\bl 1\preceq\bar{\bl u}\preceq z_{\max}\bl
1\nonumber
\end{eqnarray}
\end{enumerate}
Thus, the optimality in Lemma \ref{lem_cond_grad} still holds in
fluid limits.
\end{lemma}
We first outline the proof. For 1), since \textsf{GRAD-XBAR} an LP
over a simplex, the optimum must correspond to the vertices with the
smallest gradient. Thus, it is sufficient to prove that any resource
allocation mode $j$ will have $\bar{v}^{(m)}_j(t)=0$ if there is a
$k$ such that
\begin{equation}
(\nabla_m\bar{g}(t))_j>(\nabla_m\bar{g}(t))_k
\end{equation}
which follows from the optimality shown in Lemma \ref{lem_cond_grad}
along a convergent subsequence. For 2), we will prove that for any
feasible points $\bar{\bl u}$ of (\ref{eqn_lp_u}), we have
\begin{equation}
\nabla_{\bl z}\bar{g}(t)^T\bar{\bl u}(t)\leq \nabla_{\bl
z}\bar{g}(t)^T\bar{\bl u}
\end{equation}
which also follows from the optimality in Lemma \ref{lem_cond_grad}
along a convergent subsequence.

\begin{IEEEproof}
For the clarity of presentation, the proof is moved to the Appendix.
\end{IEEEproof}

Based on the above lemma, we are now ready to prove that Algorithm
\ref{alg} achieves the optimal cost in the fluid limit.

\begin{lemma}\label{lem_optimal_cost}
For any fluid limit, we have for all $t>0$,
\begin{equation}
g(\bar{\bl x}(t),\bar{\bl z}(t); \bar{\bl y}(t))=g^\star
\end{equation}
where $g^\star=f(\bl x^\star_{\textsf{p}}; \bl y)+\beta p(\bl
x^\star_{\textsf{p}},\bl z^\star_{\textsf{p}}; \bl y)$. Thus, the
optimality is achieved in the fluid limit.
\end{lemma}

We first outline the proof. Note that it is always true that
\begin{equation}
g(\bar{\bl x}(t),\bar{\bl z}(t); \bar{\bl y}(t))\geq g^\star
\end{equation}
since $(\bar{\bl x}(t),\bar{\bl z}(t))$ are always feasible points
of \textsf{PEN}. Thus, the claim holds if we can prove the reverse
direction. This can be done by defining a proper ``Lyapunov''
function
\begin{equation}\label{eqn_lyapunov}
L(t)=tg(\bar{\bl x}(t),\bar{\bl z}(t); \bar{\bl y}(t))
\end{equation}
and show that $\dot{L}(t)\leq g^\star$, by using the properties in
Lemma \ref{lem_fluid_grad} and the convexity of function $g(\cdot)$.

\begin{IEEEproof}
Consider the ``Lyapunov'' function as in (\ref{eqn_lyapunov}) in any
fluid limit. From Lemma \ref{lem_feasible} we know that for any
$t>0$, $(\bar{\bl x}(t),\bar{\bl z}(t))$ are feasible for
\textsf{PEN}, and therefore we have $L(t)\geq g^\star t$ due to the
definition of $g^\star$. On the other hand,
\begin{eqnarray*}
\dot{L}(t)&=&\bar{g}(t)+t\dot{\bar{g}}({t})\\
&=&\bar{g}(t)+t\nabla_{\bl z}\bar{g}(t)^T\dot{\bar{\bl
z}}(t)+t\sum_{m\in\ml M}\nabla_m\bar{g}(t)^T\dot{\bar{\bl
x}}^{(m)}(t)\\
&\stackrel{(a)}{=}&\bar{g}(t)+t\nabla_{\bl z}\bar{g}(t)^T(\bar{\bl
u}(t)-{\bar{\bl
z}}(t))\\
&&{+}\: \sum_{m\in\ml M}\nabla_m\bar{g}(t)^T({\bar{\bl
v}}^{(m)}(t)-{\bar{\bl x}}^{(m)}(t))\\
&\stackrel{(b)}{\leq}&\bar{g}(t)+t\nabla_{\bl z}\bar{g}(t)^T({\bl
z}^\star_{\textsf{p}}-{\bar{\bl z}}(t))\\
&&{+}\:\sum_{m\in\ml M}\nabla_m\bar{g}(t)^T({{\bl
x}}^{(m)\star}_{\textsf{p}}-{\bar{\bl x}}^{(m)}(t))\\
&\stackrel{(c)}{\leq}&{g}({\bl x}^\star_{\textsf{p}}, {\bl
z}^\star_{\textsf{p}})\stackrel{(d)}{=}g^\star
\end{eqnarray*}
where $(a)$ is obtained by substituting the equation in Lemma
\ref{lem_feasible}, $(b)$ is from Lemma \ref{lem_fluid_grad}, i.e.,
$\bar{\bl v}^{(m)}(t)$ and $\bar{\bl u}(t)$ are solutions of
\textsf{GRAD-XBAR} and \textsf{GRAD-ZBAR}, respectively. $(c)$ is
due to the convexity of function $g(\cdot)$, and $(d)$ is because
$({\bl x}^\star_{\textsf{p}}, {\bl z}^\star_{\textsf{p}})$ is the
solution of \textsf{PEN}, by definition. Thus, we have
\begin{equation}
L(t)=L(0)+\int_0^t \dot{L}(\tau)d\tau\leq g^\star t
\end{equation}
from which we conclude that $L(t)=g^\star t$.
\end{IEEEproof}

Having established the optimality in the fluid limit, we are now
able to prove optimality in the original system. The following
theorem states that Algorithm \ref{alg} achieves the optimal cost in
the original network.
\begin{theorem}(\emph{optimal cost})\label{the_cost}
In the original network, the following holds w.p.1:
\begin{equation}
\limsup_{t\rightarrow\infty}f(\bl x(t); \bl y(t))\leq
f_\epsilon^\star
\end{equation}
\end{theorem}
\begin{IEEEproof}
Suppose that it is not true. Then there is a sequence
$\{t_n\}_{n=1}^\infty$ such that
\begin{equation}
\lim_{n\rightarrow\infty}f({\bl x}(t_n); \bl
y(t_n))>f_\epsilon^\star
\end{equation}
From the Arzela-Ascoli Theorem \cite{royden}, there is a subsequence
$\{t_{n_k}\}_{k=1}^\infty$ which converges to a fluid limit. Lemma
\ref{lem_optimal_cost} implies
\begin{eqnarray}
\lim_{k\rightarrow\infty}f({\bl
x}(t_{n_k}))&\leq&\lim_{k\rightarrow\infty}g({\bl
x}(t_{n_k}))\\
&\stackrel{(a)}{=}&g(\bar{\bl x}(1))\\
&\stackrel{(b)}{=}&g^\star\leq f_\epsilon^\star
\end{eqnarray}
where $(a)$ follows from the fact that for all $m\in\ml M$,
\begin{eqnarray}
\bl x^{(m)}(t_{n_k})&=&\bl T^{(m)}(t_{n_k})/\bl 1^T \bl
T^{(m)}(t_{n_k})\\
&=&(\bl T^{(m)})^{t_{n_k}}(1)/\bl 1^T (\bl T^{(m)})^{t_{n_k}}(1)\\
&\rightarrow&\bar{\bl T}^{(m)}(1)/\bl 1^T\bar{\bl T}^{(m)}(1)\
\textrm{as } k\rightarrow\infty\\
&=&\bar{\bl x}^{(m)}(1)
\end{eqnarray}
and that $g(\cdot)$ is continuous. $(b)$ is because of Lemma
\ref{lem_optimal_cost}. Thus, we have a contradiction, and the claim
holds.
\end{IEEEproof}

In the next subsection we will continue to prove the feasibility
result, namely, the limit points of $\bl x(t)$ produced by Algorithm
\ref{alg} are indeed feasible for \textsf{OPT}.

\subsection{Feasibility Proof}

Note that Algorithm \ref{alg} is designed to solve \textsf{PEN}.
Thus, in order to prove that the scheduler produce feasible points
for \textsf{OPT}, we need the following lemma, which connects the
objective function value in \textsf{PEN} to the constraint in
\textsf{OPT}.

\begin{lemma}\label{lem_opt1}
The following properties hold for \textsf{PEN}: For large enough
$\beta$, we have
\begin{equation}\label{eqn_equal_const_optimal}
\|\bl h({\bl x}^\star_{\textsf{p}}; \bl y)+{\bl
z}^\star_{\textsf{p}}\|\leq \epsilon/2
\end{equation}
for any solution $({\bl x}^\star_{\textsf{p}}, \bl
z^\star_{\textsf{p}})$.
\end{lemma}
\begin{IEEEproof}
For the ease of presentation, we only consider the penalty function
as (\ref{eqn_penalty_example}), although the proof can be easily
extended to general cases. Note that from Lemma \ref{lem_approx_2}
we have
\begin{equation}
{\beta\over\alpha}\|\bl h({\bl x}^\star_{\textsf{p}}; \bl y)+{\bl
z}^\star_{\textsf{p}}\|^\alpha\leq f_\epsilon^\star-f({\bl
x}^\star_{\textsf{p}}; \bl y)
\end{equation}
Thus, (\ref{eqn_equal_const_optimal}) holds by choosing sufficiently
large $\beta$.
\end{IEEEproof}

Finally, we conclude this section by the following theorem, which
states that the limit points produced by Algorithm \ref{alg} are
always feasible for the original problem \textsf{OPT}. This,
combined with Theorem \ref{the_cost}, proves the optimality of
Algorithm \ref{alg} for \textsf{OPT}.

\begin{theorem}(\emph{feasibility})
For sufficiently large $\beta$, we have
\begin{equation}
\limsup_{t\rightarrow\infty} h_i(\bl x(t); \bl y(t))\leq 0
\end{equation}
for any constraint function $h_i(\bl x; \bl y)$ in $\bl h(\bl x; \bl
y)$.
\end{theorem}

\begin{IEEEproof}
Suppose that this is not true. Then there exist a sequence
$\{t_n\}_{n=1}^\infty$ such that
\begin{equation}\label{eqn_posi_h}
\lim_{n\rightarrow\infty}h_i({\bl x}(t_n); \bl y(t_n))>0
\end{equation}
From Arzela-Ascoli Theorem, there is a subsequence
$\{t_{n_k}\}_{k=1}^\infty$ which converges to a fluid limit. Thus,
we have
\begin{eqnarray*}
\|\bl h(\bar{\bl x}(1); \bar{\bl y}(1))+\bar{\bl
z}(1)\|&\geq&h_i(\bar{\bl
x}(1); \bar{\bl y}(1))+\bar{z}_i(1)\\
&\stackrel{(a)}{=}&\lim_{k\rightarrow\infty}h_i({\bl x}(t_{n_k});
{\bl
y}(t_{n_k}))+{z}_i(t_{n_k})\\
&\stackrel{(b)}{\geq}& \lim_{k\rightarrow\infty}h_i({\bl
x}(t_{n_k}); {\bl
y}(t_{n_k}))+\epsilon\\
&\stackrel{(c)}{\geq}& \epsilon
\end{eqnarray*}
where $(a)$ can be argued similarly as in the proof of Theorem
\ref{the_cost}, $(b)$ is because for any $i$ and $t>0$ we have
$z_i(t)>\epsilon$, due to Algorithm \ref{alg}, and $(c)$ is because
of the assumption in (\ref{eqn_posi_h}). But according to Lemma
\ref{lem_optimal_cost}, $(\bar{\bl x}(t)), \bar{\bl z}(t))$ solves
\textsf{PEN}, and therefore Lemma \ref{lem_opt1} implies that
\begin{equation}
\|\bl h(\bar{\bl x}(1); \bar{\bl y}(1))+\bar{\bl z}(1)\|\leq
\epsilon/2
\end{equation}
Contradiction! Therefore the claim holds.
\end{IEEEproof}
Thus, Algorithm \ref{alg} produces feasible points for \textsf{OPT},
and achieves a cost which is arbitrarily close to $f^\star$, by
properly selecting parameters $\beta$ and $\epsilon$.

\section{Simulation}
\label{sec_simulation}

\begin{figure}
\centering
\includegraphics[width=3.4in]{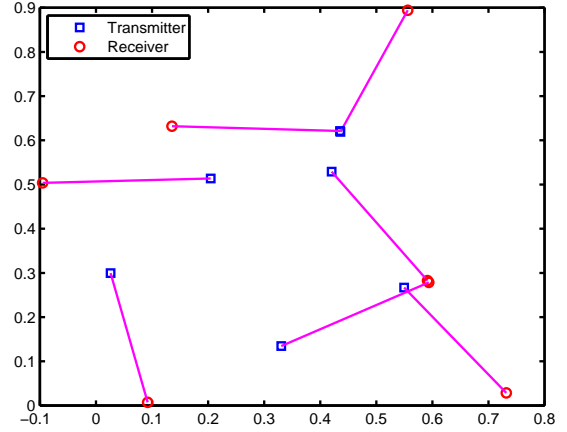}
\vspace{-0.1in} \caption{An example wireless network with 7 links,
where square nodes are transmitters, and round nodes are receivers.
} \label{fig_network}
\end{figure}

In this section we verify the performance of Algorithm \ref{alg}
through a simulation in a random wireless network where the network
is as shown in Fig .\ref{fig_network}. There are 7 links in the
network, where square nodes denote the transmitters, and round nodes
denote the receivers. We simulate a special case of \textsf{OPT},
the following minimum power scheduling problem, which we denote as
\textsf{POW}:
\begin{eqnarray}
\textsf{POW: }\min_{{\boldsymbol x}}&&\bl p^T \bl x\\
\textrm{s.t.}&&\bl a-G\bl x\preceq{\bl 0}\label{eqn_rate_stab}\\
&&{\boldsymbol x}\succeq 0,\ {\boldsymbol 1}^T{\boldsymbol x}=1
\end{eqnarray}
where $\bl p$ is a power vector whose each element $p_k$ corresponds
to the power consumption when the independent set $\bl G_k$ is
chosen. In the simulation we choose $p_k=\|\bl G_k\|^2$. Here, $\bl
a$ corresponds to the arrival rate vector, which is assumed to be
the only unknown parameter in the network. Thus,
(\ref{eqn_rate_stab}) corresponds to the rate stability constraint.

\begin{figure}
\centering
\includegraphics[width=3.6in]{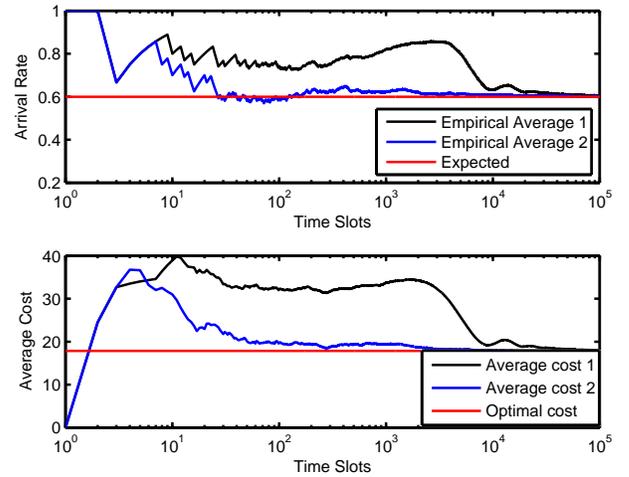}
\vspace{-0.1in} \caption{Convergence results of average cost with
sources with different convergence behaviors.} \label{fig_utility}
\end{figure}

Fig. \ref{fig_utility} shows the convergence results of the cost
function (the bottom sub-figure) with slowly converging sources (the
top sub-figure) after a simulation of $10^5$ time slots. In the
simulation, we choose $\epsilon=10^{-3}$ and $\beta=5\times10^3$. It
can be observed from the top sub-figure that our algorithm achieves
the optimal cost. Further, by comparing the convergence results of
the cost and the arrival processes, we can conclude that Algorithm
\ref{alg} can track the uncertain parameter $\bl a$ dynamically. In
the simulation, it is further observed that the maximum queue length
in the network is around $10^2$, so that the constraint in
(\ref{eqn_rate_stab}) is clearly satisfied.

\section{Conclusion}
\label{sec_conclusion}

In this paper we formulated a general class of scheduling problems
in wireless networks with uncertain parameters, subject to the
constraint that these parameters can be obtained from the empirical
average values of certain stochastic network processes. We proposed
a class of primal-dual type scheduling algorithms, and showed its
optimality as well as feasibility using fluid limits.

\appendix[Proof of Lemma \ref{lem_fluid_grad}]
\begin{IEEEproof}
We first prove 1). Let a sequence of functions $({\bl T}^{r_n}, {\bl
Y}^{r_n}, {\bl Z}^{r_n})$ be given, which converge to a fluid limit
$(\bar{\bl T}, \bar{\bl Y}, \bar{\bl Z})$. In the fluid limit,
suppose that there is time $t>0$, $m\in\ml M$ and resource
allocation modes $j, k$ such that
\begin{equation}
(\nabla_m \bar{g}(t))_j\geq (\nabla_m \bar{g}(t))_k+\epsilon
\end{equation}
where $\epsilon>0$ is a small constant. Then, since $\nabla_m
\bar{g}(t)$ is a continuous function of variable $t$, there is
$\delta_1>0$ such that for all $\tau\in(t-\delta_1, t+\delta_1)$, we
have
\begin{equation}
(\nabla_m \bar{g}(\tau))_j\geq (\nabla_m \bar{g}(\tau))_k+\epsilon/2
\end{equation}
Further, since $\nabla_m \bar{g}(\tau)\triangleq\nabla_m g(\bar{\bl
x}(\tau), \bar{\bl z}(\tau); \bar{\bl y}(\tau)))$ is continuous as a
function of variables $(\bar{\bl x}(\tau), \bar{\bl y}(\tau),
\bar{\bl z}(\tau))$ (and therefore is absolutely continuous when
restricted to a compact local region), there is an $\epsilon'>0$
such that $\|(\bl x, \bl y, \bl z)-(\bar{\bl x}(\tau), \bar{\bl
y}(\tau), \bar{\bl z}(\tau))\|\leq \epsilon'$ implies that
\begin{eqnarray*}
\big(\nabla_{m}g\big({\bl x},{\bl z}; {\bl y})\big)_j\geq
\big(\nabla_{m}g(\bar{\bl x}(\tau), \bar{\bl z}(\tau); \bar{\bl
y}(\tau))\big)_k+\epsilon/4
\end{eqnarray*}
for all $\tau\in(t-\delta_1, t+\delta_1)$. Now we define
\begin{eqnarray}
({\bl x}^{r_{n}})^{(m)}(\tau)&\triangleq&\bl T^{(m)}(r_n\tau)/\bl
1^T\bl
T^{(m)}(r_n\tau)\label{eqn_x_rn}\\
{\bl y}^{r_{n}}(\tau)&\triangleq&\bl Y(r_n\tau)/r_n\tau\label{eqn_y_rn}\\
{\bl z}^{r_{n}}(\tau)&\triangleq&\bl
Z(r_n\tau)/r_n\tau\label{eqn_z_rn}
\end{eqnarray}
Then, the definition of fluid limits implies that there exists
$N\in\mathbb N$ and $\delta_2>0$ such that for all $n>N$ and
$\tau\in(t-\delta_2, t+\delta_2)$,
\begin{eqnarray}\label{eqn_rn_bar}
\|({{\bl x}^{r_{n}}(\tau)},{{\bl y}}^{r_n}(\tau), {{\bl
z}}^{r_n}(\tau))-(\bar{\bl x}(\tau), \bar{\bl y}(\tau), \bar{\bl
z}(\tau))\|<\epsilon'
\end{eqnarray}
Thus, by taking $\delta=\min(\delta_1, \delta_2)$ we have
\begin{eqnarray*}
\big(\nabla_{m}g\big({{\bl x}^{r_{n}}(\tau)},{{\bl
z}}^{r_n}(\tau); {{\bl y}}^{r_n}(\tau)\big)_j\\
\geq \big(\nabla_{m}g({{\bl x}^{r_{n}}(\tau)},{{\bl z}}^{r_n}(\tau);
{{\bl y}}^{r_n}(\tau)\big)_k+\epsilon/4
\end{eqnarray*}
for all $\tau\in(t-\delta, t+\delta)$. Further, by comparing the
above definitions of ${{\bl x}^{r_{n}}(\tau)},{{\bl y}}^{r_n}(\tau)$
and  ${{\bl z}}^{r_n}(\tau)$ to that of $\bl x(\tau)$, $\bl y(\tau)$
and $\bl z(\tau)$ in (\ref{eqn_x}), (\ref{eqn_y}) and (\ref{eqn_z}),
respectively, we conclude that they are essentially the same, except
a difference in time scale, i.e., $\bl x^{r_n}(t)=\bl x(r_nt)$.
Thus, the following holds in the original system: for any $n>N$ and
all $\tau\in(r_n(t-\delta), r_n(t+\delta))$,
\begin{eqnarray*}
\big(\nabla_{m}g({{\bl x}(\tau)},{{\bl z}}(\tau); {{\bl
y}}(\tau)\big)_j\geq \big(\nabla_{m}g({{\bl x}(\tau)},{{\bl
z}}(\tau); {{\bl y}}(\tau)\big)_k+\epsilon/4
\end{eqnarray*}
Therefore, according to Lemma \ref{lem_cond_grad}, $\xi^{(m)}_j$ is
never chosen in any time slot during $(r_n(t-\delta),
r_n(t+\delta))$, and we have that $T_j^{(m)}(\tau)$ is a constant
during $(r_n(t-\delta), r_n(t+\delta))$, from which we conclude that
$\dot{\bar{T}}_j^{(m)}(t)=0$. Therefore, $\bar{v}^{(m)}_j(t)=0$
following the definition that
$\bar{v}_j^{(m)}(t)=\dot{\bar{T}}_j^{(m)}(t)/\pi^{(m)}$.

We next prove 2). Let $\bar{\bl u}$ be given as a feasible point of
\textsf{GRAD-ZBAR} and $\epsilon>0$ be given. Since $\nabla_{\bl z}
g(\cdot)$ is a continuous function of $t$, there is $\delta>0$ and
$N\in\mathbb N$ such that for $n>N$ and all $\tau<(t-\delta,
t+\delta)$, the following holds:
\begin{equation}\label{eqn_norm_diff}
\|\nabla_{\bl z}g({{\bl x}^{r_{n}}(\tau)},{{\bl z}}^{r_n}(\tau);
{{\bl y}}^{r_n}(\tau))-\nabla_{\bl z}{g}(\bar{\bl x}(t), \bar{\bl
z}(t); \bar{\bl y}(t))\|<\epsilon
\end{equation}
Further, note that Lemma \ref{lem_cond_grad} implies that for any
time slot in $(r_n(t-\delta), r_n(t+\delta))$, we have
\begin{eqnarray*}
\nabla_{\bl z}g({{\bl x}(\tau)},{{\bl z}(\tau)}; \bl y(\tau))^T\bl
u(\tau)\leq \nabla_{\bl z}g({\bl x}(\tau),{\bl z}(\tau); \bl
y(\tau))^T\bar{\bl u}
\end{eqnarray*}
Thus, applying (\ref{eqn_norm_diff}) to the above inequality we have
\begin{equation*}
\nabla_{\bl z}{g}(\bar{\bl x}(t), \bar{\bl z}(t); \bar{\bl
y}(t))^T\bl u(\tau)\leq\nabla_{\bl z}{g}(\bar{\bl x}(t), \bar{\bl
z}(t); \bar{\bl y}(t))^T\bar{\bl u}+c\epsilon
\end{equation*}
for all $\tau\in(r_n(t-\delta), r_n(t+\delta))$, where $c>0$ is a
proper constant. After summing over $(r_n(t-\delta), r_n(t+\delta))$
and dividing by $r_n$ on both sides, we obtain
\begin{eqnarray*}
\nabla_{\bl z}{g}(\bar{\bl x}(t), \bar{\bl z}(t); \bar{\bl
y}(t))^T(\bl Z^{r_n}(t+\delta)-\bl
Z^{r_n}(t-\delta))\\
\leq2\delta(\nabla_{\bl z}{g}(\bar{\bl x}(t), \bar{\bl z}(t);
\bar{\bl y}(t))^T\bar{\bl u}+c\epsilon)
\end{eqnarray*}
Finally, we let $n\rightarrow\infty$, and noting that $\delta>0$ can
be taken arbitrarily small, we have
\begin{equation*}
\nabla_{\bl z}{g}(\bar{\bl x}(t), \bar{\bl z}(t); \bar{\bl
y}(t))^T\bar{\bl u}(t)\leq\nabla_{\bl z}{g}(\bar{\bl x}(t),\bar{\bl
z}(t); \bar{\bl y}(t))^T\bar{\bl u}+c\epsilon
\end{equation*}
from which 2) holds since $\epsilon>0$ is arbitrary.
\end{IEEEproof}

\end{document}